# Quantum Machine Learning for Secondary Frequency Control

Younes Ghazagh Jahed[1], Alireza Khatiri[2]

[1]Department of Electrical Engineering, Faculty of Engineering and Technology, University of Mazandaran, Babolsar, Iran, y.ghazagh09@umail.umz.ac.ir, ORCID: 0009-0009-8323-3224

[2]Department of Electrical Engineering, Faculty of Engineering and Technology, University of Mazandaran, Babolsar, Iran, a.khatiri02@umail.umz.ac.ir, ORCID: 0009-0005-7321-168X

**Abstract**

Frequency control in power systems is critical to maintaining stability and preventing blackouts. Traditional methods like meta-heuristic algorithms and machine learning face limitations in real-time applicability and scalability. This paper introduces a novel approach using a pure variational quantum circuit (VQC) for real-time secondary frequency control in diesel generators. Unlike hybrid classical–quantum models, the proposed VQC operates independently during execution, eliminating latency from classical–quantum data exchange. The VQC is trained via supervised learning to map historical frequency deviations to optimal Proportional-Integral (PI) controller parameters using a pre-computed lookup table. Simulations demonstrate that the VQC achieves high prediction accuracy (over 90%) with sufficient quantum measurement shots and generalizes well across diverse test events. The quantum-optimized PI parameters significantly improve transient response, reducing frequency fluctuations and settling time.

**Keywords,** Frequency Control, Power System, Quantum Machine Learning, Quantum Neural Networks, Supervised Learning, Variational Quantum Circuits.

## Introduction

The problem of controlling the frequency of a system is a vital and crucial matter in power systems. The uncontrolled power unit, specifically generators as the hearts of power systems, can cause serious issues and might lead to cascaded failures which eventually reach the unwanted point of system blackout[1]. In this regard, to address frequency control, several researchers have developed various types of approaches: 1) Meta-heuristic algorithms are one of the most popular and classic approaches, which can be utilized in different formats, whether for obtaining optimal static parameters of a controller or for adaptively tuning controller parameters in model-predictive-based methods[2-4]. However. despite their advantages in solving optimization problems, meta-heuristic algorithms suffer from their slow and iterative nature, which makes them unsuitable for real-time applications[5]. 2) Machine learning methods are the newest and most promising solutions in recent years[6-8]. Different types of these approaches are used to directly solve optimization problems, e.g., reinforcement learning algorithms, which might not be optimal as the dimension and complexity of the problem increase. Other types of machine learning methods are utilized either for predicting the future or for real-time decision-making in a supervised learning setup. Despite the benefits of machine learning algorithms—i.e., real-time processing and handling big data—maintaining their performance for more complex problems often requires increasing the size of the model, e.g., total layers and units in neural networks[9]. Recent studies show that quantum machine learning can be a solution for handling complex data. Due to the existence of superposition and entanglement in quantum mechanics, variational quantum circuits (VQCs) may extract better features from complex data compared with classical algorithms[10]. It is noteworthy that this advantage is observed when comparing quantum neural networks and classical neural networks with the same size (i.e., the same total number of trainable parameters)[11]. VQCs have been utilized in hybrid classical–quantum formats for frequency control; however, the delay caused by processing and exchanging signals between the classical computer with its CPU and the quantum computer with its QPU might cause issues in real-time operations.

This paper attempts to develop a pure VQC without any aid from classical computers during real-time operations. To the best of the authors' knowledge, this is the first attempt to utilize a purely quantum circuit for tuning frequency controller parameters.

## Secondary Frequency Control in Diesel Generators

In power system operation, maintaining frequency stability is crucial for ensuring power quality and system reliability. For diesel generators, secondary frequency control complements the primary droop control to eliminate steady-state frequency errors following load variations. The control structure comprises a cascade of dynamic elements that collectively regulate the generator's active power output.

The control loop begins with a Proportional-Integral (PI) controller, whose transfer function is represented as[12]:

$$K(s) = K_p + \frac{K_i}{s} \quad (1)$$

where $K_p$ and $K_i$ denote the proportional and integral gains, respectively. The output of this controller, denoted as $\Delta P_{\text{ref}}$, is combined with the primary control action derived from the frequency deviation $\Delta f$ through the droop coefficient $R$. The resulting signal becomes[12]:

$$\Delta P_c = \Delta P_{\text{ref}} - \frac{\Delta f}{R} \quad (2)$$

This composite signal $\Delta P_c$ serves as the input to the governor system, which exhibits first-order dynamics characterized by the time constant $T_G$ [12]:

$$G_G(s) = \frac{1}{T_G s + 1} \quad (3)$$

The governor output then drives the diesel engine actuator, modelled with its own time constant $T_{\text{DG}}$ [12]:

$$G_{\text{DG}}(s) = \frac{1}{T_{\text{DG}} s + 1} \quad (4)$$

The mechanical power output $\Delta P_m$ from the prime mover is subsequently subjected to load disturbances $\Delta P_L$. The net power imbalance $(\Delta P_m - \Delta P_L)$ enters the generator's inertia model, which incorporates the combined effects of rotational inertia $H$ and load damping $D$ [12]:

$$G_I(s) = \frac{1}{Hs + D} \quad (5)$$

The output of this final block yields the system frequency deviation $\Delta f$, thereby closing the control loop. The integral action in the PI controller ensures asymptotic rejection of frequency errors, while the droop component provides proportional power sharing during transients. This hierarchical control structure enables precise frequency regulation while maintaining stable operation across varying load conditions.

## Variational Quantum Circuit for Supervised Parameter Learning

This work presents a VQC approach for supervised learning of optimal proportional-integral controller parameters from historical frequency deviation data. The quantum circuit operates within the framework of quantum supervised learning, where the objective is to learn a mapping from input frequency deviations $[\Delta f_{t-k}, \ldots, \Delta f_{t-2}, \Delta f_{t-1}]$ to optimal PI parameters obtained from a pre-computed look-up table.

The quantum computational framework begins by encoding classical input data into quantum states. For an input vector $x = [\Delta f_{t-k}, \ldots, \Delta f_{t-2}, \Delta f_{t-1}] \in \mathbb{R}^k$, the state preparation circuit maps this classical information onto a three-qubit quantum register. The initial quantum state is prepared as $|0\rangle \ldots |0\rangle$, and through a series of unitary transformations, the input features are encoded into the quantum state $|\psi_{in}\rangle$.

The VQC comprises two primary components: the feature map layer and the parameterized ansatz. The feature map implements a nonlinear transformation of the input data into a high-dimensional Hilbert space, facilitating the expression of complex relationships between input features. This transformation is achieved through the application of Hadamard gates followed by parameterized phase rotations. The feature map state can be represented as:

$$|\psi_{fm}\rangle = U_{fm}(x) |\overbrace{0 \ldots 0}^{k}\rangle \quad (6)$$

where $U_{fm}(x)$ represents the unitary operation that encodes the input features $x$ into the quantum state through rotations and entangling gates.

The parameterized ansatz $U(\theta)$ follows the feature map and consists of alternating layers of single-qubit rotations and entangling gates. The ansatz structure implements a unitary transformation parameterized by the trainable parameters $\theta$:

$$U(\theta) = \prod_{layer=1}^{L} U_{ent} \cdot U_{rot}(\theta_{layer}) \quad (7)$$

where $U_{rot}(\theta_{layer})$ applies single-qubit rotations and $U_{ent}$ creates entanglement between qubits using controlled-NOT gates.

The complete quantum state evolution can be described by the sequence:

$$|\psi(x,\theta)\rangle = U(\theta)U_{fm}(x)|\overbrace{0\ldots 0}^{k}\rangle \quad (8)$$

where the initial state $|0\ldots 0\rangle$ evolves through the feature map $U_{fm}(x)$ and then through the parameterized ansatz $U(\theta)$.

The quantum circuit's output is obtained through measurement in the computational basis. For each input sample, the circuit is executed multiple times to estimate the expectation values of the measurement operators. The measurement process yields a probability distribution over the computational basis states:

$$P(b) = |\langle b|\psi(x,\theta)\rangle|^2 \quad (9)$$

where $b$ denotes the binary measurement outcomes from the set {000,001,010,011,100,101,110,111}.

The objective function for this optimization is formulated as the mean squared error between the circuit predictions and the target PI parameters from the look-up table:

$$\mathcal{L}(\theta) = \frac{1}{N}\sum_{i=1}^{N}\|y_i - f(x_i;\theta)\|^2 \quad (10)$$

where $N$ represents the number of training samples, $y_i$ denotes the target PI parameters, and $f(x_i;\theta)$ represents the quantum circuit's output for input $x_i$ with parameters $\theta$.

The optimization algorithm iteratively updates the parameter vector $\theta$ by moving candidate solutions toward the best solution. The trained VQC thus learns to approximate the functional relationship between historical frequency deviations and optimal PI controller parameters (replay memory), leveraging quantum superposition and entanglement to capture nonlinear dependencies.

The first step in training process focuses on data generation and collection. We employ a standard power system model, which includes components such as the Governor and the Diesel Generator block (representing the dynamics of the system being controlled), along with their inherent inertia. The system's behaviour is managed by a conventional Controller block, K(s), which parameters are continuously tuned by a Optimizer. This Optimizer functions as an expert agent, systematically searching for and implementing the control actions that yield the best system performance. The control inputs and the resulting system state information—specifically, sequences of optimal frequency deviations, represented as $[\Delta f_{t-3}, \Delta f_{t-2}, \Delta f_{t-1}]$, and the corresponding optimal control output Y generated by the expert system—are stored sequentially in a buffer called the Replay Memory. This memory builds the high-quality, optimal dataset necessary for the subsequent supervised training. The second step transitions to the supervised training of the VQC. Data pairs are extracted from the Replay Memory, where a sequence of historical frequency deviations serves as the input and the corresponding optimal control signal Y (from the expert) serves as the desired target output. The VQC itself is implemented using three quantum bits (qubits) and comprises alternating layers of parameterized rotation gates (indicated by blocks like P, RX, and RZ) and two-qubit entangling gates (marked with a plus sign, which represent CNOT-like operations). These gates allow the circuit to process information and generate a control output, Y'. The VQC's calculated output, Y', is then evaluated against the expert's target output, Y, using a Loss Function. This function quantifies the difference between the quantum circuit's prediction and the desired expert action. Finally, a second Optimizer is employed to minimize this loss. It iteratively adjusts the internal, adjustable parameters of the VQC's rotation gates, effectively teaching the VQC to replicate the expert's optimal control decisions. Through this process, the VQC is trained to perform as an optimal controller, harnessing the power of quantum computing for enhanced control performance.

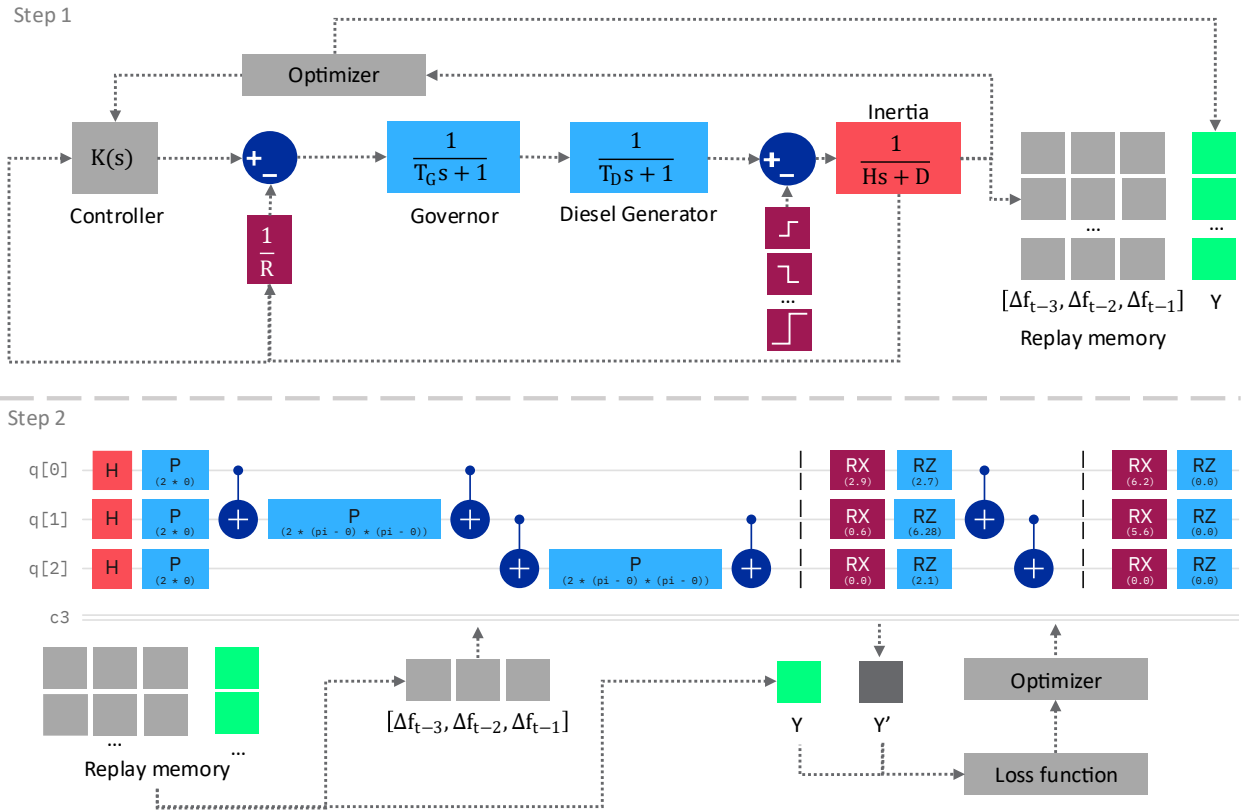

**Figure 1. Training Process of the Quantum Control System.** The training of our novel quantum control architecture follows a two-step supervised learning process. This approach leverages the performance of an optimized classical controller, which acts as a domain expert, to train an VQC to mimic its optimal control strategy.

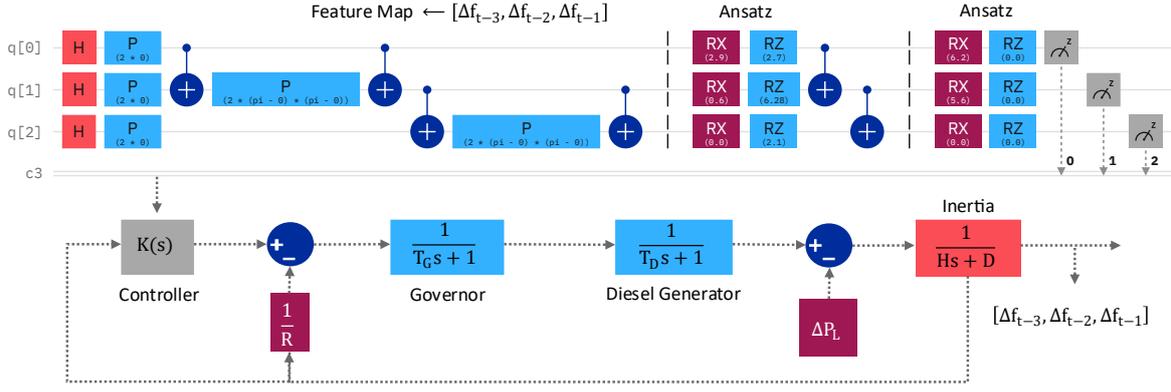

**Figure 2. Operation Stage of the Hybrid Quantum Control System.** Following the supervised training process, the optimized VQC is deployed to replace the classical controller K(s) and operate the system in real time. This stage represents the execution of the learned control policy.

In the operation stage, the control system functions as a closed-loop feedback mechanism. The system dynamics, which include the Governor and the Diesel Generator blocks, respond to changes in load ($\Delta P_L$) and system inertia. The objective is to maintain system frequency stability. Input Preparation: The latest sequence of historical frequency deviations, represented by the vector $[\Delta f_{t-3}, \Delta f_{t-2}, \Delta f_{t-1}]$, is continuously sampled from the system output. This sequence serves as the input state for the quantum controller. Feature Mapping (Encoding): The input data is first processed by the VQC's initial layers, collectively known as the *Feature Map*. This map consists of fixed gates (Hadamard gates *H* and parameterized *P* gates with fixed arguments) and entangling gates. The Feature Map's role is crucial: it translates the classical input signals into a high-dimensional quantum state, which is required for quantum computation. Ansatz (Computation): The encoded quantum state then passes through the remaining layers of the VQC, referred to as the *Ansatz*. This section contains the learned, optimized parameters (the numerical values shown on the *RX* and *RZ* rotation gates) resulting from the supervised training. The Ansatz performs the core computational task, acting as the control policy by manipulating the quantum state to determine the appropriate control action. Output (Control Signal Generation): The final quantum state is measured (indicated by the measurement gates at the end of the circuit). The measured expectation values from the qubits $q[0], q[1], q[2]$ are processed to generate the control signal, $K(s)$ (or the equivalent output control effort). Control Action and Feedback: This generated control signal is injected into the system to counteract disturbances and regulate the frequency. The resulting frequency deviation is fed back to the input preparation stage, establishing the continuous closed-loop operation of the hybrid quantum-classical control system. The VQC, having been successfully trained on optimal expert data, performs the complex control task by leveraging its quantum processing capabilities, effectively replacing the traditional classical controller in this final deployment configuration.

# Result

All simulations in this study were implemented using the Python programming language, leveraging the IBM Qiskit library for constructing and executing quantum circuits. The computational experiments were conducted in the Google Colab environment, which provides a cloud-based platform with access to necessary computational resources, including GPU and CPU support. The figure 3 presents a statistical analysis of the VQC's accuracy across different resource allocations, specifically varying the number of times the quantum circuit is run and measured (the "Shots configuration"). This directly addresses the robustness and scalability of the quantum approach. X-Axis (Shots configuration): Represents the number of repetitions used for quantum measurements: 100, 1000, 5000, and 10000 shots. Increasing the number of shots generally reduces shot noise (measurement variance). Y-Axis (Accuracy): Represents the classification accuracy of the VQC on the test set, ranging from 0.6 to 1.0 (60% to 100%). The large square/pentagonal marker (connected by the thick vertical line) represents the mean accuracy for that specific shots' configuration. The visualization clearly shows a strong correlation between the number of measurement shots and the stability/performance of the VQC. In Low Shots (100), the mean accuracy is the lowest (around 0.80), The variance is the highest. This large spread is characteristic of high shot noise. When few shots are used, the statistical fluctuations in the measurement outcomes significantly impact the final classification, leading to inconsistent accuracy. As the number of shots increases, the performance stabilizes and improves. The mean accuracy quickly jumps to around 0.93 - 0.94 and remains consistently high. Crucially, the variance (error bar length) drastically shrinks. At 5000 and 10000 shots, the range of observed accuracies is significantly smaller, with the minimum, mean, and maximum all tightly clustered near 1.0.

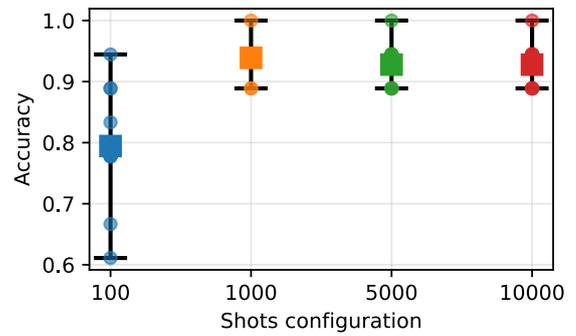

**Figure 3. Impact of Measurement Shots Configuration on VQC Prediction Accuracy.** Increasing the number of shots to 1000 or more effectively mitigates quantum measurement noise, leading to a stable, high-performance model (average accuracy above 90%) with minimal variance.

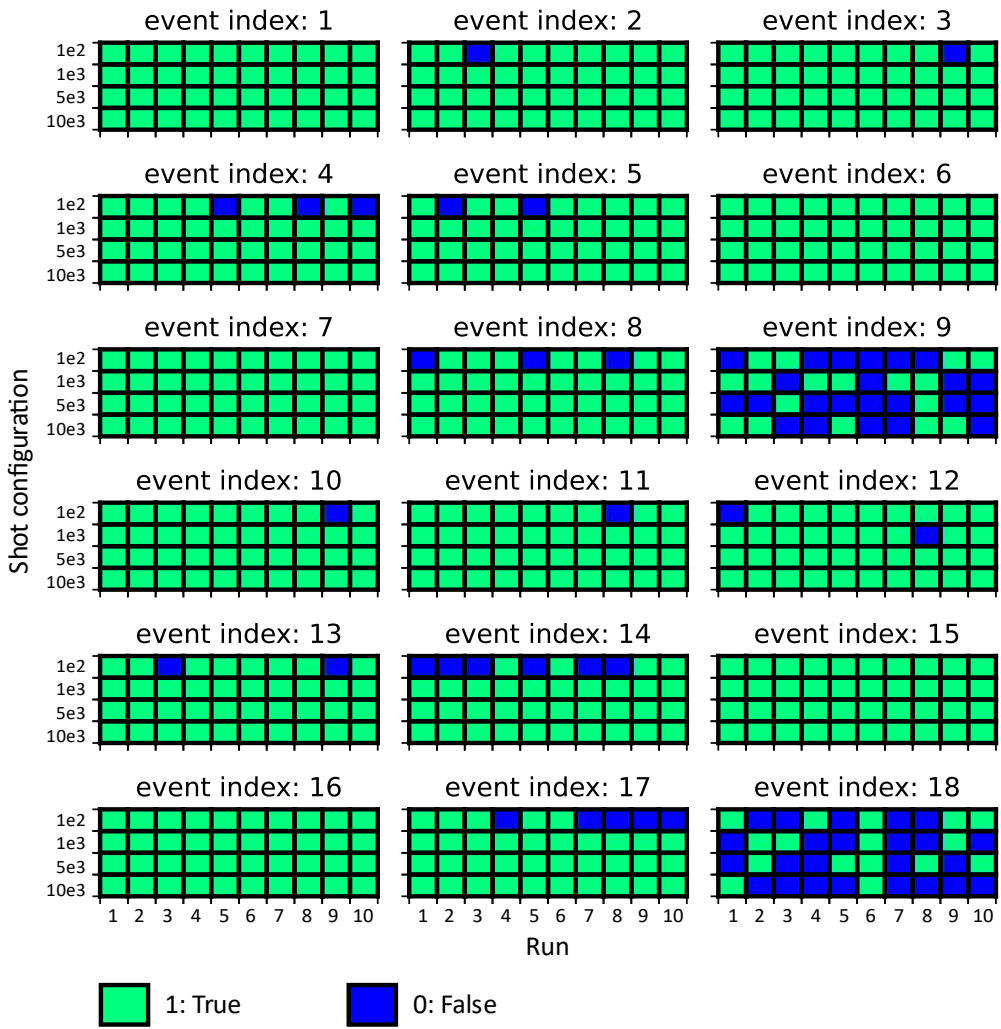

**Figure 4. Validation of Quantum Neural Network Performance Across Test Events.** The generalization capability of the VQC model on the test set is confirmed. While the model achieves perfect or near-perfect accuracy for the majority of the test events, the observed "False" predictions in certain events (the blue cells) demonstrate areas where the VQC's output is sensitive to the run-to-run variation and/or the specific measurement settings ("Shot configuration"). The goal is to maximize the green area across all 18 events, proving the robustness and reliability of the quantum supervised learning approach for control system parameter optimization.

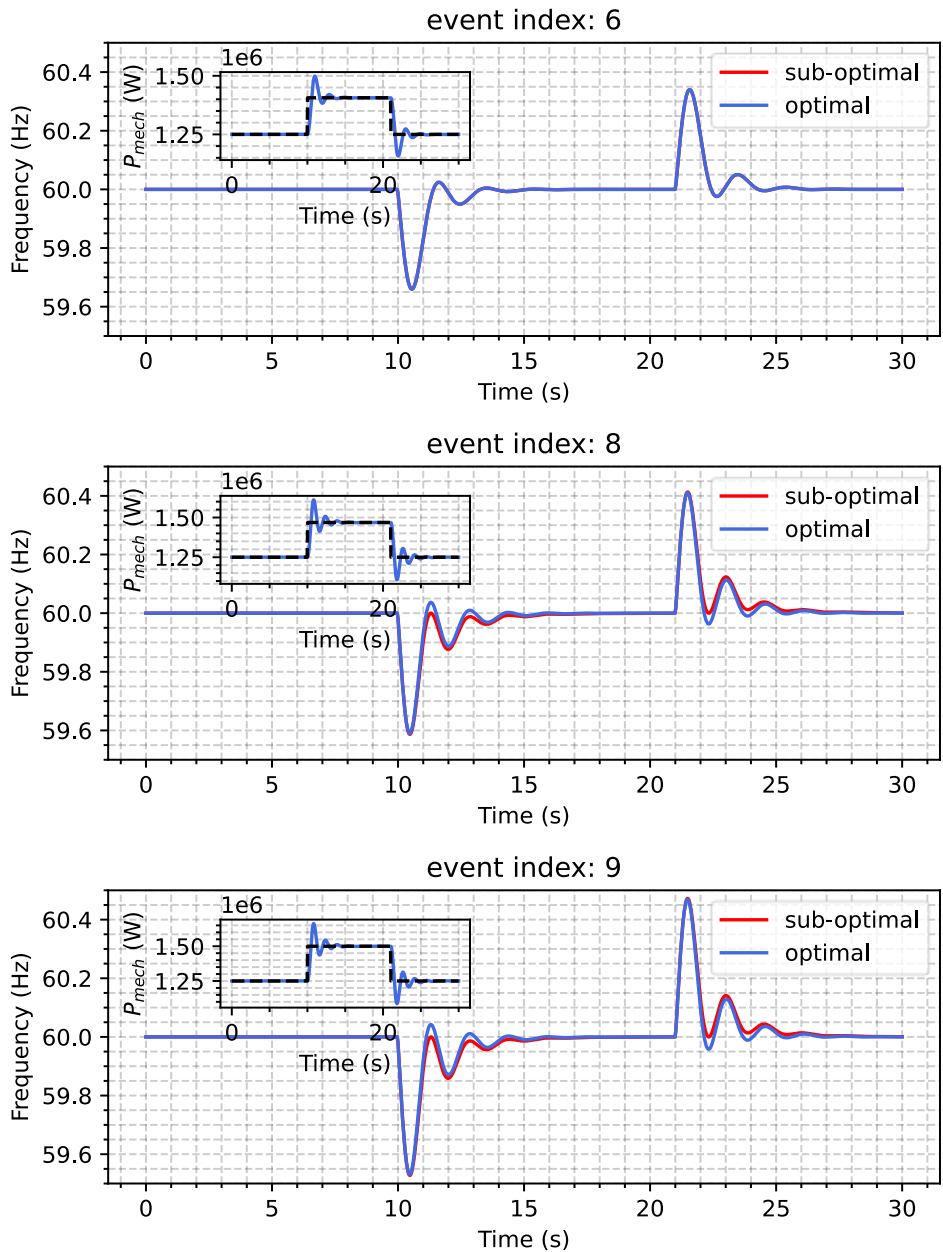

**Figure 5. Power System Frequency Response.** Time-domain simulation results for the system frequency following load disturbances at 10s and 21s, shown for three test events. The blue line (optimal) uses the PI parameters predicted by the VQC model. The red line (sub-optimal) uses a false VQC prediction. The optimal parameters consistently yield better dynamic performance, characterized by a higher frequency nadir (less severe drop) and faster damping/settling time. The inset shows the mechanical power response, confirming the optimal controller's superior transient tracking capability.

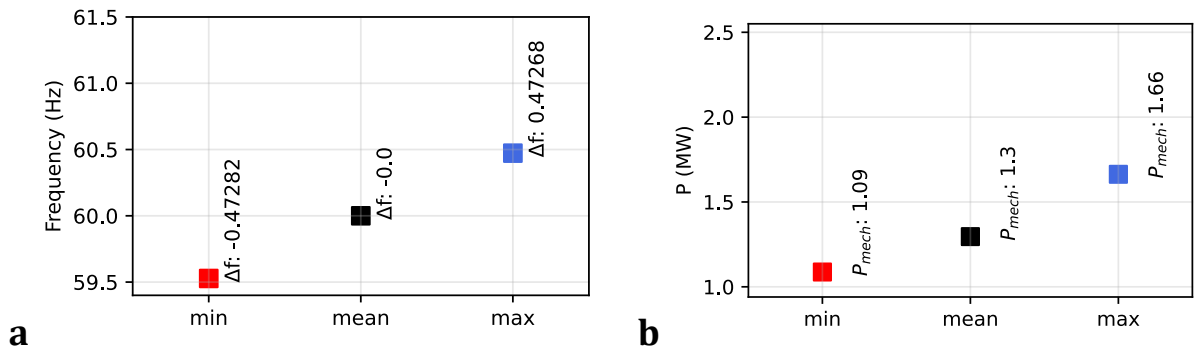

**Figure 6. Statistical Performance Summary of the Power System Regulated by VQC-Optimized PI Parameters.** (a) Statistical Summary of System Frequency Deviation. The plot illustrates the distribution of the system frequency across all test events. (b) Statistical Summary of Mechanical Power Output. This plot quantifies the control effort exerted by the generator across the entire test set. The markers show the minimum (1.09), mean (1.3), and maximum (1.66) mechanical power output values observed.

Figure 4 displays the prediction accuracy of the trained VQC model across a test set comprising 18 distinct events (labelled as "event index: 1" through "event index: 18"). The visualization is presented as an array of 18 heatmaps, one for each test event. The Vertical Axis (y-axis) is implicitly labelled "Shot configuration" which represents different measurement settings. The Horizontal Axis (x-axis) is labelled "Run," representing different instances of the measurement process for a given event and shot configuration (Runs 1 through 10). Green (1: True): Indicates that the VQC model's predicted set of PI controller parameters for that specific "Run" and "Shot configuration" matches the optimal PI parameters obtained from the pre-computed look-up table (the ground truth). Blue (0: False): Indicates that the VQC model's predicted set of PI controller parameters does not match the optimal parameters (a misclassification or sub-optimal prediction). The performance is assessed by observing the proportion of Green (True) cells for each event. High Accuracy Events (e.g., Event Index 1, 2, 3, 10, 12, 16): For these events, the grid is almost entirely Green. This signifies that the VQC model consistently and accurately mapped the input frequency deviation to the correct optimal PI parameters, regardless of the "Shot configuration" or "Run" instance. This indicates the model has learned the underlying pattern for these specific test cases very well. Lower Accuracy/Inconsistent Events (e.g., Event Index 9, 15, 18): These events show a significant number of Blue (False) cells, particularly in the later Runs (e.g., Runs 6-10) and potentially varying across Shot configurations. The presence of blue cells means the VQC failed to predict the optimal PI parameters in those instances.

The simulation results conclusively demonstrate that the optimal PI controller parameters learned by the VQC model provide significantly superior transient stability and dynamic performance for the power system compared to the sub-optimal parameters. This superiority is evidenced by reduced frequency nadir, improved damping, and shorter settling time following disturbances, thereby validating the utility and effectiveness of the quantum machine learning approach for critical control system applications.

This figure 6(a) provides a concise statistical summary of the power system frequency deviation observed across the entire test set of events. The plot displays three key statistical markers on the y-axis, which represents the system frequency in Hz: the minimum (min) frequency reached during any test event, the mean (average) frequency across all events, and the maximum (max) frequency reached. The perfect mean frequency of 60.0 Hz confirms the overall stability and effective regulation capacity of the control system utilizing the optimal PI parameters derived from the VQC approach. Figure 6(b) provides a statistical summary of the mechanical power output of the generator across the entire test set of events.

## Conclusions

This study successfully developed and validated a pure VQC for real-time frequency control in power systems. The VQC was trained to predict optimal PI controller parameters using historical frequency data, achieving high accuracy and robust performance across multiple test scenarios. Key findings include: The VQC's performance stabilizes at high accuracy (above 90%) with increased quantum measurement shots (≥1000), mitigating shot noise effects. The model generalizes well across most test events, though some variability persists in complex cases, indicating sensitivity to circuit depth and measurement settings. The fully quantum approach eliminates classical–quantum hybrid delays, making it suitable for real-time control. These results underscore the feasibility of quantum machine learning for critical power system applications. Future work will focus on scaling the VQC for larger systems, improving robustness under noise, and exploring real-hardware deployment.

**Declaration of competing interest**

The authors declare that they have no known competing financial interests or personal relationships that could have appeared to influence the work reported in this paper.